\documentclass[conference]{IEEEtran}
\IEEEoverridecommandlockouts

\usepackage{acronym}
\usepackage{booktabs}
\usepackage{amsmath,amssymb,amsfonts}
\usepackage{algorithmic}
\usepackage{array}
\usepackage{cite}
\usepackage{comment}
\usepackage{graphicx}
\usepackage{multirow}
\usepackage{textcomp}
\usepackage{xcolor}
\usepackage{booktabs}
\def\BibTeX{{\rm B\kern-.05em{\sc i\kern-.025em b}\kern-.08em
    T\kern-.1667em\lower.7ex\hbox{E}\kern-.125emX}}

\usepackage{hyperref}
\hypersetup{
    colorlinks=true,
    citecolor=green,
    filecolor=black,
    linkcolor=red,
    urlcolor=blue
}
\usepackage{cleveref}

\acrodef{acpf}[ACPF]{alternating current power flow}
\acrodef{acopf}[ACOPF]{alternating current optimal power flow}
\acrodef{ai}[AI]{artificial intelligence}
\acrodef{scacopf}[SC ACOPF]{security-constrained alternating current optimal power flow}
\acrodef{cpu}[CPU]{central processing unit}
\acrodefplural{cpu}[CPUs]{central processing units}
\acrodef{gpu}[GPU]{graphical processing unit}
\acrodefplural{gpu}[GPUs]{graphical processing units}
\acrodef{hpc}[HPC]{high-performance computing}
\acrodef{llm}[LLM]{large language model}
\acrodef{rag}[RAG]{retrieval augmented generation}
\acrodef{rto}[RTO]{regional transmission operator}
\acrodef{simd}[SIMD]{single-instruction multiple-data}
\acrodef{uc}[UC]{unit commitment}
\acrodefplural{uc}[UCs]{unit commitments}
\acrodef{dduc}[DDUC]{data-driven unit commitment }
\acrodef{sc}[S.C.]{South Carolina}

\begin{document}

\title{Agentic Artificial Intelligence for Power Systems: Strategies to Identify and Close Capability Gaps%
\thanks{This manuscript has been authored by UT-Battelle, LLC under Contract No.\ DE-AC05-00OR22725 with the U.S.\ Department of Energy. The publisher acknowledges the US government license to provide public access under the DOE Public Access Plan (http://energy.gov/downloads/doe-publicaccess-plan).}
}

\author{
\IEEEauthorblockN{Eve Tsybina}
\IEEEauthorblockA{\textit{Oak Ridge National Laboratory} \\
Oak Ridge, TN, USA \\
tsybinae@ornl.gov}
\and
\IEEEauthorblockN{Samim Konjicija}
\IEEEauthorblockA{\textit{University of Sarajevo} \\
Sarajevo, Bosnia and Herzegovina \\
skonjicija@etf.unsa.ba}
\and
\IEEEauthorblockN{Slaven Peles}
\IEEEauthorblockA{\textit{Oak Ridge National Laboratory} \\
Oak Ridge, TN, USA \\
peless@ornl.gov}
}

\maketitle

\begin{abstract}
The rapid expansion of AI-driven information infrastructure, particularly data centers, is placing unprecedented pressure on power systems and accelerating the pace at which new assets must interconnect with the grid. As bulk transmission expansion rolls out slowly, new loads and generation are increasingly deployed within existing network constraints. Agentic AI is urgently needed to automate the numerous and repetitive connection processes, but its maturity has not been systematically validated on complex tasks and large-scale systems. We replicate the current state of the art in agentic AI for power systems planning and evaluate it against a structured suite of nodal planning problems spanning six levels of task complexity and four grid scales. We find that only the two lowest complexity levels are solvable on some of the test grid sizes, and identify the specific capability upgrades required to close this gap. Adopting stricter testing protocols and reproducible evaluation benchmarks is essential for assessing both genuine progress and the operational readiness of agentic AI.
\end{abstract}

\begin{IEEEkeywords}
agentic artificial intelligence, grid planning.
\end{IEEEkeywords}

\section{Introduction}
AI-driven information economy, particularly data center expansion, is placing significant pressure on power systems. According to Q4 2025 S\&P Global Market Intelligence report \cite{b1}, data centers have added 43.8~GW of utility load in the United States, 1.3~GW in Canada, and 7.8~GW in Europe over the past five years. An additional 114.2~GW in the U.S.\ are planned for deployment during 2026--2030, 16.2~GW in the EU and 3.7~GW in Canada. This rapid load growth is outpacing expansion in generation and transmission, leading to a more congested and more localized grid. New load pockets emerge, and part of the grid that previously operated as a single area may split into multiple constrained subareas. While grid expansion or reinforcements are frequently discussed, they remain a slow, long-term process. Experience from regions with coordinated transmission planning, such as the EU and China \cite{entso2026tyndp,christoffersen2018china}, shows that planning and construction of bulk grid transmission infrastructure takes decades. In the context of the AI race and in decentralized investment environments like the U.S., investors are unlikely to wait for grid upgrades. Instead, development proceeds within existing constraints, adding load or ``own'' generation where possible, deferring asset retirements, and relying on minimal network modifications \cite{b2,b3}.

The practical baseline is therefore that new assets -- loads, generation, or grid-support devices -- must connect within large-scale grids while assuming the grid as constant and relying on nodal studies to find feasible locations. This requires agentic AI to operate at both high resolution and interconnection scale grids. Such operation encompasses a broad class of engineering problems, which can be well represented by, for example, a connection study, or a local substation upgrade study. There is a growing body of research on agentic AI rising to the challenge \cite{zhang2025poweragent, zhang2025grid, jin2025gridmind, b_llm4opf, cheng2025gaia}. However, it is unknown at this point whether existing level of agentic development can support the necessary planning processes. As we are adopting agentic AI to copilot high performance power system software
, we attempted to (a) understand the existing agentic AI efforts, as they are discussed in publicly available research, and (b) assess the extent to which agentic AI can solve nodal large scale planning problems. Based on published results, we developed our own AI agent prototype, assessed its performance, and identified immediate next steps for improvement.  

The resulting manuscript is structured as follows. Section~\ref{sec:survey} provides a survey of existing agentic AI literature. Section~\ref{sec:methodology} discusses the experimental methodology for agentic AI profiling. Section~\ref{sec:results} discusses the efforts taken to replicate the state of the art agentic performance, and presents test results. Section~\ref{sec:conclusions} summarizes our findings and outlines directions for future research.

\section{State of the Art Survey}
\label{sec:survey}


We focus on agentic AI products, known as "copilots" that can operate with both, physics-based and AI inference models. Other applications of AI include generative models for visualization \cite{chaturvedi2025chatgrid} and foundational models aimed to replace state estimators or physics-based simulation \cite{choi2024generative,cheng2026ai}. For more information about different uses of \ac{llm} and generative AI in power systems see reviews in \cite{amjad2025review,mirshekali2025review,b_mdpi}. Perhaps the best description of architecture and workflow fundamentals of \ac{llm} instruments for power systems can be found in \cite{b_llm4opf}. 

It is hard to provide a comprehensive literature review in a fast-growing research space, as new products appear every day. There is an increasing number of conceptual research papers and frameworks. Simultaneously, few studies report the results of deploying actual agents, and even fewer disclose code for verification. As of June~2026, we identified five grid-specific agentic AI solutions (Table \ref{tab:agents}).

\begin{table*}[hbt]
\caption{Comparison of Grid-Specific Agentic AI Solutions}
\label{tab:agents}
\begin{center}
\begin{tabular}{|>{\raggedright\arraybackslash}m{3cm}|>{\raggedright\arraybackslash}m{5cm}|>{\raggedright\arraybackslash}m{3cm}|>{\raggedright\arraybackslash}m{3cm}|>{\raggedright\arraybackslash}m{1.9cm}|}
\hline
\textbf{Agent} & \textbf{Use Cases} & \textbf{Vendor Models} & \textbf{Test System} & \textbf{Code, data available} \\
\hline
PowerAgent\newline (US, Harvard) 2025 \cite{zhang2025poweragent} &
Deterministic OPF;\newline N-1 contingencies: identify harmful contingencies;\newline Load growth analysis & Claude &
4-, 7-, and 34-bus systems 
(exact source unknown)
 &
Code only \\
\hline
Grid-Agent\newline (Canada) 2025 \cite{zhang2025grid} &
Contingency management through battery usage, load shed, or switch operation.\newline
Has continuous learning capability.
 & gemini-2.5-pro, gemini-2.5-flash, gemini-2.5-flash-lite, gpt-4.1, gpt-4.1-mini, gpt-4.1-nano &
CIGRE MV (14-bus), IEEE 30, 69* &
No \\
\hline
GridMind\newline (US, Argonne) 2025 \cite{jin2025gridmind} &
Deterministic OPF;\newline
N-1 contingencies: identify harmful contingencies & GPT-5, GPT-5-mini, GPT-5-nano, GPT-o3, GPT-o4-mini, Claude 4 Sonnet &
IEEE 14, 30, 118, 300\cite{ieee118bus}** &
No \\
\hline
LLM4OPF\newline (Australia, Hong Kong) 2023 \cite{b_llm4opf} &
EV charging control\newline
The Git repo contains extra applications not specified in the document & GPT-4 and GPT-4 Vision &
Single bus system &
Yes \\
\hline
GAIA\newline (China, Hong Kong, Singapore) 2025 \cite{cheng2025gaia} &
Grid monitoring and operation, blackstart\newline
Does not interface with solvers
 & LLaMA2, GPT-3.5, GPT-4 &
IEEE 14, 30, 57, 118* &
No \\
\hline

\end{tabular}
\end{center}
* There are multiple versions of IEEE n-bus systems, some of them deviating significantly from the original one. The exact source and therefore the exact configuration of the test systems are unknown.

** The IEEE systems contained in \cite{ieee118bus} are "copperplate" with infinite transmission and unlimited MW and MVar (+/-) generation. Unless explicitly modified, their behavior is close to a single bus system.  
\end{table*}

Based on the available information, all of the discussed agentic AI tools share the same fundamental elements:
\begin{itemize}
    \item storage for registering power system state;
    \item storage with \ac{rag} helpers that allow the AI to understand industry language and inform specific tasks;
    \item solver interface that feeds data into power flow models such as \ac{acopf};
    \item executor module that issues commands to adjust the state and run the simulations;
    \item context module that saves results of earlier iterations to inform new iterations;
    \item validator and summary module that interprets results.
\end{itemize}

The only exception is \cite{cheng2025gaia}, which did not go as far as executing commands. All the referenced AI agents are reported to be capable of solving basic problems on small grids. Examples of problems include:
\begin{itemize}
    \item ``Solve IEEE test case'' \cite{jin2025gridmind};
    \item ``Increase load in bus 10 to 150~MW'' \cite{jin2025gridmind};
    \item ``Increase load to 101\%'' \cite{zhang2025poweragent}.
\end{itemize}

More complex tasks such as resolving contingency by grid switching reveal some problems in developing and adjusting respective agents. For instance, GridAgent \cite{zhang2025grid} reports that a separate agent is needed for each separate type of contingency resolution. PowerAgent \cite{zhang2025poweragent} similarly reports generalization issues. This is also found in non-agentic, generative AI research studies that deal with linguistic interpretation, and general \ac{llm} studies investigating \ac{rag} usage\cite{choi2024generative,yan2024linguistic,cheng2025gaia}. 

System size and resulting token cost is another specific problem found in agentic studies\cite{zhang2025poweragent}. A close analysis of agentic solutions indicates that the problem is probably partially attributed to the use of JSON files. JSON format is used by all major AI vendors to send and receive data. It uses deep nesting, repeating keys, parses slowly, and is not memory efficient compared to other input types. A large JSON file can take hundreds of thousands of tokens to convey information to the \ac{llm}. It was also shown \cite{zhang2025grid} that large systems exceed the context capacity of vendor LLMs, making it impossible for the model to retain a description of the entire system. Often, ``healthy'' parts of the system are retained while in the model's context while the parts requiring attention are dropped.

Unfortunately, little is reported about the scaling of the agentic AI performance with the size of the power system problem and the change in agentic iteration requirements with the complexity of prompts. 
While \cite{jin2025gridmind} reports the absence of correlation between search time and system size, this finding may be misleading. There, authors use IEEE cases from University of Washington archive \cite{ieee118bus}, which are ``copperplate'' infinite transmission, infinite generation systems. Therefore an increase in the number of buses may not necessarily lead to an increase in the number of agent iterations or a significant increase in solution time. Since none of the referenced studies provides source data for the test cases, it is hard to tell if the adopted systems were closer to ``copperplate'' systems, or to realistic congestible systems. The latter would require more iterations and more time to find the feasible solution.

The absence of clear verifiable code is the main barrier to understanding the genuine ability of the proposed agents to perform the required power system simulations. We attempt to replicate the state of the art code as discussed in the next section, and test it against a more demanding set of tasks and grid scales.

\section{Experimental Setup}
\label{sec:methodology}

\subsection{Analysis parameters}

Assessing the operational readiness of agentic AI is difficult because few research results have been documented in sufficient detail. Nevertheless, the results published so far suggest that the community is converging on the common design principles discussed in \cref{sec:survey}. This trend allows us to construct a ``generic'' agent and evaluate its performance on tasks of increasing complexity. Our agent is based on a development snapshot of AgentiGrid, an AI agent integrated with the \ac{hpc} power flow analysis package ExaGO \cite{peles2026exago}. It incorporates all of the agentic AI elements discussed in \cref{sec:survey} and closely reflects the state of the art described in the literature. The source code is available at \cite{peles2026exago}
.

Next, we design the test cases. A comprehensive assessment of agentic AI
across the full range of connection study tasks would require simulations
spanning many design choices, including the \ac{llm} vendor, the grid size,
and the prompt design. Together, these choices span a prohibitively large
experimental space. As a first approximation, we therefore restrict our study
to the following design:
\begin{align*}
  2 \text{ vendors} &\times 6 \text{ prompts} \times 4 \text{ grid sizes}
  \times 10 \text{ attempts}=\\ &= 480 \text{ simulations.}    
\end{align*}

We select two vendors from different regions, DeepSeek and OpenAI, leaving
other vendors such as Anthropic (Claude) and Google (Gemini) to future work.
The second design choice is grid size. Since the largest systems reported in
the literature are on the order of 100--200 buses, and system size may affect
AI performance, we test four sizes: 100 buses (the state-of-the-art average),
1{,}000 buses (a mid-sized utility), 10{,}000 buses (a large transmission
operator), and 100{,}000 buses (a stretch target representing a combined
transmission and distribution system). We further use six prompts of
increasing complexity, described in the next section. Finally, because
\ac{llm} responses are not deterministic -- repeating the same prompt may yield
different answers 
-- we allow 10 attempts per prompt to
obtain a consistent answer. Each simulation records the AI's search history,
prompt response, token usage, analytics, and processing time.

\subsection{Experimental Grid and Scenarios}

To test the scalability of agentic AI for grid planning, we built grids of
adjustable size by tiling copies of a 100-bus building block. The block is
derived from the IEEE 118-bus three-area system \cite{ieee118bus}, reduced to
100 buses to allow even size increments, and then adjusted, following the PNNL
manual \cite{PNNL}, to introduce transmission and generating-unit constraints.
This adjustment is critical for testing: operating a grid that congests and
has reactive power issues is disproportionately more complex than operating a
``copperplate'' grid. The adjusted grid (Fig.~\ref{fig:adjusted_topology}) is
provided in the supplemental material.

\begin{figure}[h]
\centerline{\includegraphics[width=\columnwidth]{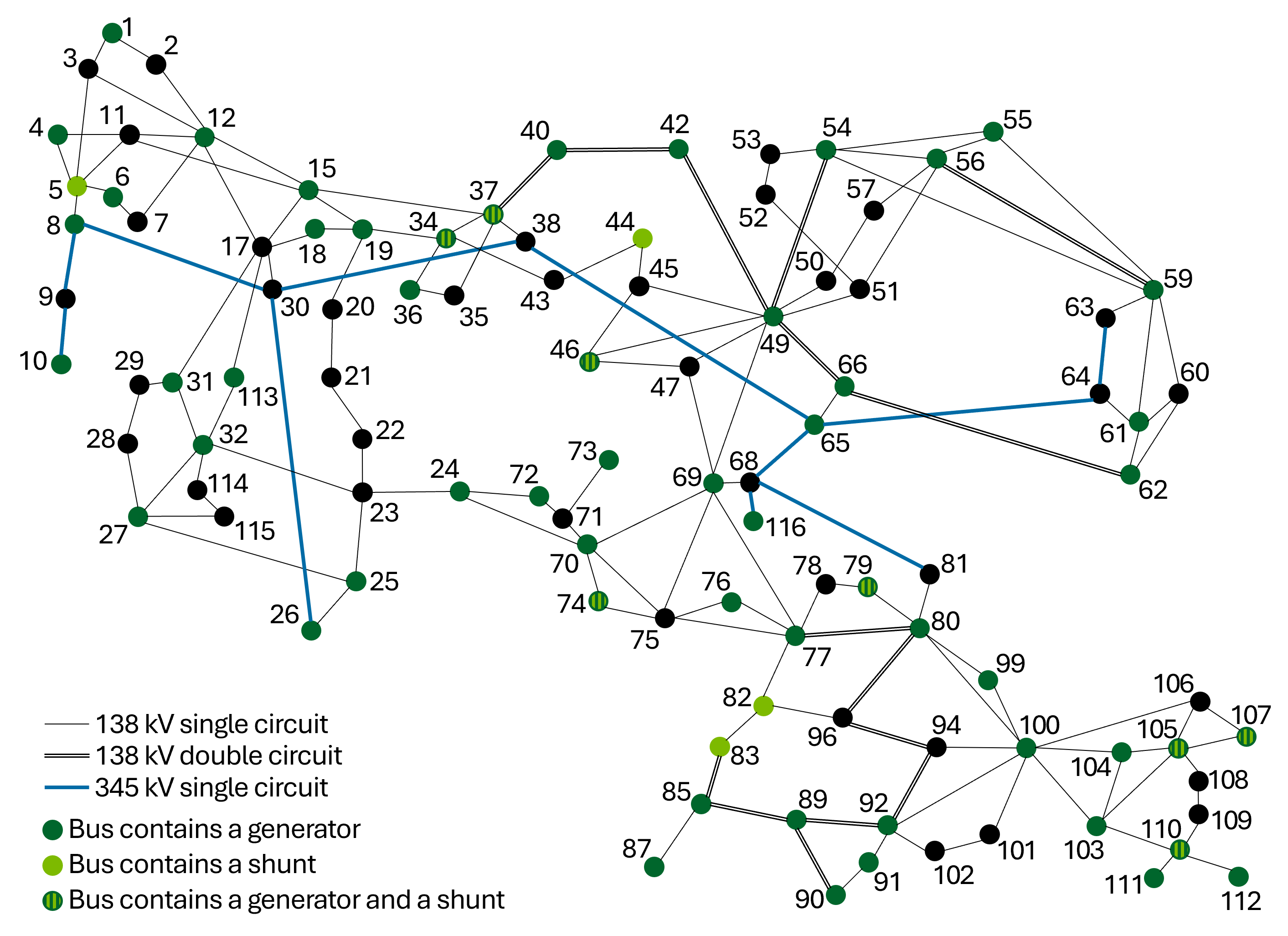}}
\caption{IEEE 118-bus grid adjusted for profiling.}
\label{fig:adjusted_topology}
\end{figure}

Since the objective of this study is to test the ability of agentic AI to
perform load and generation studies, we evaluate the agent with a set of
prompts addressing optimal siting or hosting capacity, together with two key
steady-state outcomes of a generation and load portfolio: grid congestion and
hot reserve. Each prompt is assigned a difficulty level from 0 to 6, defined as the number
of discrete operations the agent must perform to get from the base case to the
final result. We call these operations hops. Each hop's output is consumed by
the next. Hops include constructing a candidate set (e.g., enumerating
contingencies or identifying nearest neighbors), solving a case, evaluating
feasibility, aggregating results (e.g., min/max/average), and selecting or
filtering. For example, computing the system reserve margin from nameplate
data is a single aggregation hop (Level 1); testing N-1 contingencies requires
constructing, solving, and evaluating the single-element contingency set
(Level 3); and testing N-2 contingencies adds two further hops to construct
and solve the pairwise set (Level 5). Most prompts from
\cite{jin2025gridmind} require no hops at all (Level 0).

[Level 1] Please find all buses that can host a 100 MW, 10 MVar load / using transmission constrained ACOPF / assume that the feasibility criteria include: (1) bus voltage magnitude in p.u. between 0.9 and 1.1, (2) line MVA flow is within rate A limit.

[Level 2] Please report the reactive adequacy of the system for a 100 MW, 50 Mvar generator connected to each bus in the system / using transmission constrained ACOPF / assume that reactive adequacy holds if a generator is capable of providing reactive power at $Q_{\max}$ while providing active power at $P_{\max}$.

[Level 3] Please test all N-1 contingencies on 3 nearest neighbor buses after connecting a 50 MW load to bus $\langle$bus number$\rangle$, indicate the passed and failed contingencies / contingencies include branch outage, generator outage, and load outage / assume nearest neighbors by hop count / assume that the feasibility criteria include: (1) bus voltage magnitude in p.u. between 0.9 and 1.1, (2) line MVA flow is within rate A limit.

[Level 4] Please find the minimum feasible hot reserve in the system for N-1 contingencies / assuming that hot reserve is the difference between maximum MW of the units that have status “on” and scheduled MW output of the units that have status “on” / assume that the feasibility criteria include: (1) bus voltage magnitude in p.u. between 0.9 and 1.1, (2) line MVA flow is within rate A limit.

[Level 5] Please test all N-2 contingencies on 3 nearest neighbor buses after connecting a 50 MW load to bus $\langle$bus number$\rangle$, indicate the passed and failed contingencies / contingencies include branch outage, generator outage, and load outage / assume nearest neighbors by hop count / assume that the feasibility criteria include: (1) bus voltage magnitude in p.u. between 0.9 and 1.1, (2) line MVA flow is within rate A limit.

[Level 6] Please test all N-2 contingencies on 3 nearest neighbor buses after connecting a 50 MW load to bus $\langle$bus number$\rangle$ and suggest contingency relief measures / contingencies include branch outage, generator outage, and load outage / assume nearest neighbors by hop count / assume that the feasibility criteria include: (1) bus voltage magnitude in p.u. between 0.9 and 1.1, (2) line MVA flow is within rate A limit / assume the following relief measures, in the order of decreasing priority: transformer ratio change, generator redispatch, line switching, load curtailment.

We used Claude Opus 4.8 to validate the prompts before submitting them to the
agent and to aggregate the results of the agent's reasoning. Because this model comes from a vendor other than the two under study, the
prompt preprocessing and result postprocessing cannot bias the measured
performance of either model.

All the analyses were run on a workstation with Intel Core i7-11700F processor with 8 cores, and with 64 GB of RAM. 

\section{Results and Discussion}
\label{sec:results}

The overall findings of our simulations confirm the results presented in the
literature. As shown in \cref{tab:results}, our agentic AI prototype handled
prompts at the complexity level reported for existing agents. It also went
beyond prompt difficulty Level~1, the current state of practice, to Level~2,
suggesting that the industry-wide approach to storage, helpers, and interfaces
can handle both basic screening tasks and more complex analytical tasks. Both 
configurations, one using a DeepSeek and the other an OpenAI \ac{llm},
attempted the easier prompts on small grids. 
The OpenAI configuration solved both prompts in all 10 attempts; the DeepSeek
configuration solved the Level~1 prompt in 5 of 10 attempts and omitted the
10~MVar reactive load in the other 5, on both the 100-bus and 1,000-bus grids. In the
failed attempts, the agent correctly diagnosed the omission and attempted a
corrected sweep by adding the 10~MVar in the next iteration, but the cache
fingerprinted the corrected sweep as identical to the flawed original and
returned the cached results instead of re-solving. This indicates that the
sweep cache key excludes the reactive-load parameter.

\begin{table}[htbp]
  \centering
  \caption{Agentic AI performance on the state of the art grids and prompts, showing numbers of successful and total attempts.}
  \label{tab:results}
  \begin{tabular}{llrr}
    \toprule
    \multirow{2}{*}{LLM} & \multirow{2}{*}{Prompt}
      & \multicolumn{2}{c}{Grid size (buses)} \\
    \cmidrule{3-4}
      & & 100 & 1,000 \\
    \midrule
    \multirow{2}{*}{deepseek v4 pro} & [Level 1] Find all buses \ldots & 5/10 & 5/10 \\
                             & [Level 2] Report the reactive \ldots & 10/10 & 10/10 \\
    \midrule
    \multirow{2}{*}{gpt 5.4 B} & [Level 1] Find all buses \ldots & 10/10 & 10/10 \\
                             & [Level 2] Report the reactive \ldots & 10/10 & 10/10 \\
    \bottomrule
  \end{tabular}
\end{table}

As a consequence, for 100-bus grids, correct attempts by OpenAI and DeepSeek spent about 17,000 tokens, but incorrect attempts by DeepSeek spent 27,000--39,000 tokens trying to revise the original incorrect numbers, with token costs scaling with the number of iterations. For the 1,000-bus grid,
correct attempts used about 36,000--43,000 tokens, and incorrect ones between
57,000 and 78,000 tokens. Extra AI efforts, such as selecting the lowest-cost
solution and making analytical recommendations, were not consistent with the
numbers produced by the solver, for either vendor or any system. Even when the
solver sweeps were implemented correctly, the AI did not pick the correct
numbers from the solver outputs or did not reason correctly about the
solutions.

Our agentic AI prototype was unable to solve more complex prompts and larger cases. Problems exist for both larger grids and more complex prompts. While the precise causes and solutions are beyond the scope of this methodology study, we briefly report several findings that illustrate why the proposed advanced testing matters for developing better agentic AI.  

First, we examined why larger grids fail even at simple prompt levels (runs
with 10,000- and 100,000-bus grids returned no results and are omitted from
\cref{tab:results}). The bottleneck is the backend execution, not the \ac{llm}
or the agentic architecture: in the code snapshot used for our experiments,
each candidate evaluation deep-copied the network state and used linear
lookups, giving a roughly $O(N^2)$ sweep cost; in addition, the \ac{llm}
payload enumerated all network elements and overflowed the context window.
Both issues have been addressed in a more recent version of AgentiGrid, which
now responds to Level~1 and~2 prompts on the 10,000-bus grid (results
forthcoming); a memory bottleneck at 100,000 buses remains, which we expect to
resolve via batched computations on a distributed memory hardware. 

Second, more complex prompts 3-6 failed to produce results. They require a larger and more detailed set of helper functions and dictionaries: a function that iterates over buses and tests contingencies does not know how to resolve that contingency. This confirms the findings of \cite{zhang2025grid} at a more advanced testing level. The agent starts failing immediately one level of difficulty above the one reported in \cite{zhang2025poweragent, zhang2025grid}. For difficulty Level 3, the OpenAI agent used considerably more tokens - 50,000-100,000, but did not produce any results. It never sent any contingencies into the solver. The iterations focused almost exclusively on identifying what constitutes a contingency and exploring the grid for nearby buses, branches, and other grid elements. DeepSeek added a 50 MW load to bus 77, and found it feasible. In one of the attempts, it identified tier-1 nearest neighbors (buses 69, 80, 82), and listed the contingencies that would need testing. The DeepSeek attempts used 160,000-170,000 tokens and 300-400 seconds. This directly indicates that relying on the native capability of vendor AI with minimal \ac{rag} is not resulting in valid solutions. Expanding the number of helper dictionaries and functions is a dimension which cannot be addressed intensively, by the efficiency of code scripting, and has to addressed extensively, through more verbose \ac{rag}.

\section{Conclusions and Future Work}
\label{sec:conclusions}

This study offers an advanced testing methodology and demonstrates its
effectiveness in revealing capability gaps and areas for improvement in
agentic AI for power systems analysis. It introduces four grid sizes with tighter operational
constraints, and six prompt levels of greater complexity, than those used in
current best-practice testing of agentic AI. Applying these tests to the
``generic best practice'' agent exposed the need for more efficient backend
code and for extensive, detailed \ac{rag}. Because code and data for existing
agentic platforms are not publicly available, it is difficult to predict
exactly how those agents would perform under the proposed framework. However,
the framework provides immediate insights for developers of existing or new
agentic AI into likely next steps and upcoming challenges in developing their
products.

The insights from this study and the literature suggest two further
improvements to scalability and performance on complex tasks. First, the
backend could be improved to enable larger system sizes, or the JSON-based
storage and cache could be replaced altogether with a database solution such
as PostgreSQL. Second, where \ac{rag} becomes prohibitively long and complex,
moving away from vendor models in favor of a custom \ac{llm} trained
specifically for engineering purposes may be preferable.

\section*{Acknowledgment}

This manuscript has been authored by UT-Battelle, LLC under Contract No.\ DE-AC05-00OR22725 with the U.S.\ Department of Energy.

\bibliographystyle{IEEEtran}
\bibliography{references}

@techreport{b1,
  author        = "{S\&P Global Market Intelligence}",
  title         = "Datacenters and Energy Report",
  institution   = "S\&P Global Market Intelligence",
  year          = "2025",
  note          = "Subscription based."
}

@misc{b2,
  title         = "{FERC} Technical Conference Regarding the Challenge of Resource Adequacy in {RTO} and {ISO} Regions",
  howpublished  = "Federal Energy Regulatory Commission",
  month         = jun,
  year          = "2025"
}

@electronic{b3,
  title         = "Tech Firms Are Building Their Own Power Plants",
  organization  = "National Association of Manufacturers",
  url           = "https://nam.org/tech-firms-are-building-their-own-power-plants-34960/",
  month         = nov,
  year          = "2025",
  note          = "Accessed 11/14/2025"
}

@techreport{PNNL,
  title={A Real-Time Operations Manual for the IEEE 118 Bus Transmission Model},
  author={Anderson, Alexander and Kincic, Slaven and Jefferson, Brett and Mcgary, Blaine and Fallon, Corey and Ciesielski, Danielle and Wenskovitch, John and Chen, Yousu},
  year={2022},
  number = "PNNL-334996",
  institution={Pacific Northwest National Laboratory, Richland, Washington 99354}
}

@inproceedings{jin2025gridmind, 
    author = {Jin, Hongwei and Kim, Kibaek and Kwon, Jonghwan},
    title = {GridMind: LLMs-Powered Agents for Power System Analysis and Operations},
    year = {2025}, isbn = {9798400718717},
    publisher = {Association for Computing Machinery}, address = {New York, NY, USA},
    url = {https://doi.org/10.1145/3731599.3767409},
    doi = {10.1145/3731599.3767409},
    pages = {560–568},
    numpages = {9},
    location = { },
    series = {SC Workshops '25} 
}

@ARTICLE{zhang2025poweragent,
  author={Zhang, Qian and Xie, Le},
  journal={IEEE Power and Energy Magazine}, 
  title={PowerAgent: A Road Map Toward Agentic Intelligence in Power Systems: Foundation Model, Model Context Protocol, and Workflow}, 
  year={2025},
  volume={23},
  number={5},
  pages={93-101},
  doi={10.1109/MPE.2025.3579718}
}

@article{b_llm4opf,
  author        = {Huang, Chenghao and Li, Siyang and Liu, Ruohong and Wang, Hao and Chen, Yize},
  title         = {Large Foundation Models for Power Systems},
  journal       = {arXiv},
  year          = {2023},
  url           = {https://arxiv.org/abs/2312.07044}
}

@article{cheng2025gaia,
  title={A large language model for advanced power dispatch},
  author={Cheng, Yuheng and Zhao, Huan and Zhou, Xiyuan and Zhao, Junhua and Cao, Yuji and Yang, Chao and Cai, Xinlei},
  journal={Scientific Reports},
  volume={15},
  number={1},
  pages={8925},
  year={2025},
  publisher={Nature Publishing Group UK London}
}

@ARTICLE{amjad2025review,
  author={Amjad, Furqan and Korõtko, Tarmo and Rosin, Argo},
  journal={IEEE Access}, 
  title={Review of LLMs Applications in Electrical Power and Energy Systems}, 
  year={2025},
  volume={13},
  number={},
  pages={150951-150969},
  doi={10.1109/ACCESS.2025.3599922}
}

@ARTICLE{mirshekali2025review,
  author={Mirshekali, Hamid and Reza Shadi, Mohammad and Ghanadi Ladani, Fatemehsadat and Reza Shaker, Hamid},
  journal={IEEE Access}, 
  title={A Review of Large Language Models for Energy Systems: Applications, Challenges, and Future Prospects}, 
  year={2025},
  volume={13},
  number={},
  pages={163162-163188},
  doi={10.1109/ACCESS.2025.3610994}
}

@techreport{choi2024generative,
  title={Generative AI for power grid operations},
  author={Choi, Seong Lok and Jain, Rishabh and Feng, Cong and Emami, Patrick and Zhang, Hongming and Hong, Junho and Kim, Taesic and Park, SangWoo and Ding, Fei and Baggu, Murali and others},
  year={2024},
  number = "NREL/TP-5D00-91176",
  institution={National Renewable Energy Laboratory (NREL), Golden, CO (United States)},
  url = "https://docs.nrel.gov/docs/fy25osti/91176.pdf"
}

@techreport{cheng2026ai,
  title={AI for the Grid and AI on the Grid: Implications, Challenges and Opportunities},
  author={Cheng, Bo and Botterud, Audun and Levin, Todd and Nadarajah, Selvaprabu and Zhao, Dongwei and Kwon, Jonghwan},
  institution   = "Argonne National Laboratory",
  month = {February},
  url = {http://dx.doi.org/10.2139/ssrn.6315298},
  doi = {10.2139/ssrn.6315298},
  year={2026}
}

@INPROCEEDINGS{chaturvedi2025chatgrid,
  author={Chaturvedi, Sarthak and Jin, Sichen and Abhyankar, Shrirang and Thurber, Travis and Oikonomou, Kostas and Voisin, Nathalie},
  booktitle={2025 IEEE Power \& Energy Society General Meeting (PESGM)}, 
  title={Grid CoPilot: A Large Language Model (LLM) based framework for Transforming Long-term Planning Analyses}, 
  year={2025},
  volume={},
  number={},
  pages={1-5},
  doi = {10.1109/PESGM52009.2025.11225574},
  url = "https://ieeexplore.ieee.org/document/11225574"
}

@article{b_mdpi,
  author        = {Kiasari, Mahmoud and Aly, Hamed},
  title         = {Agentic Artificial Intelligence for Smart Grids: A Comprehensive Review of Autonomous, Safe, and Explainable Control Frameworks},
  journal       = "Energies",
  volume        = "19",
  number        = "3",
  year          = "2026",
  pages         = "617",
  url           = "https://www.mdpi.com/1996-1073/19/3/617"
}

@online{entso2026tyndp,
	title = {Entso-e {\textbar} Planning the future grid - {TYNDP}},
	url = {https://tyndp.entsoe.eu/},
	urldate = {2026-05-06},
	langid = {english},
}

@techreport{christoffersen2018china,
	title = {The Asian Super Grid in Northeast Asia and China's Belt and Road Initiative},
	author = {Christoffersen, Gaye},
	langid = {english},
    institution = {German Institute for International and Security Affairs},
    year = {2018},
    urlOPT = {https://www.swp-berlin.org/publications/products/projekt_papiere/Christoffersen_BCAS_2018_Northeast_Asian_Supergrid_13.pdf},
	urldate = {2026-05-06}
}

@ARTICLE{yan2024linguistic,
  author={Yan, Ziming and Xu, Yan},
  journal={IEEE Transactions on Power Systems}, 
  title={Real-Time Optimal Power Flow With Linguistic Stipulations: Integrating GPT-Agent and Deep Reinforcement Learning}, 
  year={2024},
  volume={39},
  number={2},
  pages={4747-4750},
  doi={10.1109/TPWRS.2023.3338961}
}

@misc{ieee118bus,
  title         = "{IEEE} 118-Bus Three-Area Test System",
  key  = "University of Washington Power Systems Test Case Archive",
  organization = "University of Washington",
  url           = "https://labs.ece.uw.edu/pstca/",
  urldate = {2026-05-06}
}

@article{zhang2025grid,
  title={Grid-agent: An LLM-powered multi-agent system for power grid control},
  author={Zhang, Yan and Saber, Ahmad Mohammad and Youssef, Amr and Kundur, Deepa},
  journal={arXiv preprint arXiv:2508.05702v3},
  url = {https://arxiv.org/html/2508.05702v3},
  urldate = {2026-05-06},
  year={2025}
}

@misc{ peles2026exago,
title = {ExaGO v2},
author = {Peles, Slaven and Koukpaizan, Nicholson and Alam, Maksudul and Hambrick, Joshua and Tsybina, Evgeniya},
abstractNote = {ExaGO is a high-performance computing power systems modeling suite providing models for different power flow analyses. It supports forward AC power flow, multiperiod AC and DC optimal power flow analyses, contingency analysis, as well as stochastic optimal power flow analysis. ExaGO can use HiOp and Ipopt optimization engines. It supports Matpower and PSS/E input file formats. ExaGO v2 includes code from ExaGO 1.6.0.},
doi = {10.11578/dc.20260223.2},
urlOPT = {https://doi.org/10.11578/dc.20260223.2},
howpublished = {[Computer Software] \url{https://doi.org/10.11578/dc.20260223.2}},
year = {2026},
monthOPT = {March}
}

\end{document}